\documentclass[aps,pra,twocolumn,showpacs,superscriptaddress]{revtex4}

\newcommand{\DDHi}{D$_2$H$^+$}

\usepackage{graphicx}
\usepackage{amsmath}
\usepackage{amssymb}
\usepackage{dcolumn}
\usepackage{color}

\begin{document}

\title{Energy-sensitive imaging detector applied to the dissociative recombination of \DDHi}


\author{H. Buhr}
\email[Corresponding author: ]{henrik.buhr@mpi-hd.mpg.de}
\affiliation{Faculty of Physics, Weizmann Institute of Science, 76100 Rehovot,
Israel}
\affiliation{Max-Planck-Institut f\"{u}r Kernphysik, 69121 Heidelberg, Germany}

\author{M. B. Mendes}
\author{O. Novotn\'{y}}
\affiliation{Max-Planck-Institut f\"{u}r Kernphysik, 69121 Heidelberg, Germany}
\author{D. Schwalm} \affiliation{Faculty of Physics, Weizmann
  Institute of Science, 76100 Rehovot, Israel}
\affiliation{Max-Planck-Institut f\"{u}r Kernphysik, 69121 Heidelberg, Germany} 
\author{M. H. Berg}
\author{D. Bing}
\affiliation{Max-Planck-Institut f\"{u}r Kernphysik, 69121 Heidelberg, Germany}
\author{O. Heber}
\affiliation{Faculty of Physics, Weizmann Institute of Science, 76100
  Rehovot, Israel}
\author{C.~Krantz}
\author{D.~A.~Orlov}
\affiliation{Max-Planck-Institut f\"{u}r Kernphysik, 69121 Heidelberg, Germany}
\author{M. L. Rappaport}
\affiliation{Faculty of Physics, Weizmann Institute of Science, 76100
  Rehovot, Israel}

\author{T.~Sorg}
\author{J. St\"{u}tzel}
\author{J. Varju}
\affiliation{Max-Planck-Institut f\"{u}r Kernphysik, 69121 Heidelberg, Germany}
 \author{A. Wolf}
\affiliation{Max-Planck-Institut f\"{u}r Kernphysik, 69121 Heidelberg, Germany}
\author{D. Zajfman}
\affiliation{Faculty of Physics, Weizmann Institute of Science, 76100
  Rehovot, Israel}

\date{\today}

\begin{abstract}
  We report on an energy-sensitive imaging detector for studying the
fragmentation of polyatomic molecules in the dissociative recombination of fast
molecular ions with electrons. The system is based on a large area
(10$\times$10\,cm$^2$) position-sensitive, double-sided Si-strip detector with
128 horizontal and 128 vertical strips, whose pulse height information is read 
out individually. The setup allows to uniquely identify fragment masses and is
thus capable of measuring branching ratios between different fragmentation
channels, kinetic energy releases, as well as breakup geometries, as a function
of the relative ion-electron energy. The properties of the detection system,
which has been installed at the TSR storage ring facility of the
Max-Planck Institute for Nuclear Physics in Heidelberg, is illustrated by an
investigation of the dissociative recombination of the deuterated triatomic
hydrogen cation \DDHi. A huge isotope effect is observed when comparing the 
relative branching ratio between the D$_2$+H and the HD+D channel;
the ratio 2$B$(D$_2$+H)/$B$(HD+D), which is measured to be $1.27\pm0.05$ at relative
electron-ion energies around 0~eV, is found to increase to $3.7\pm0.5$ at $\sim 5$~eV.

\end{abstract}

\pacs{34.80.Lx, 34.80.Ht, 34.80.Gs}

\maketitle


\section{\label{intro} Introduction}

Dissociative recombination (DR) with electrons is the most important loss
channel for molecular ions in cold dilute plasmas and a source of energetic
neutral atoms and excited molecules~\cite{DR-book}. Rate coefficients, branching
ratios, and the energy sharing between internal and external degrees of freedom of
the fragments are thus important ingredients in modeling these plasmas. Even
though a good understanding of the DR process of diatomic molecular ions has
been reached in recent years, theory is not yet capable of supplying most of the
information needed for these studies. Experimental investigations of the DR
process in particular of polyatomic systems are thus still indispensable to
further constrain the parameters of these models and to provide basic theory
with detailed and reliable results for crucial benchmark systems.

Considerable experimental progress in studying the DR process has been made
since the advent of heavy-ion storage rings with merging electron beam
facilities, which allow to control the inner excitation of the molecular ions
and the relative ion-electron energy precisely enough to produce data in the
parameter space of interest for, e.g., planetary atmospheres or interstellar
chemistry. Results obtained so far were vital in helping to unravel the
processes governing the DR and to clarify many important and pertinent issues.
But despite these experimental advances for particular diatomic systems,
detailed investigations of the DR of more complex ions, such as measurements of
branching ratios at relative energies other than zero, of the inner excitation
of the molecular fragments or of their dissociation kinematics, are still
hampered due to the limitations of presently available detection techniques in
identifying the DR reaction products.

We have therefore developed an {\bf E}nergy-sensitive {\bf MU}lti-strip detector
system (EMU) that is capable of recording multi-fragment events following the
DR of polyatomic molecular ions and to identify the fragments by their masses.
The system has been installed at the ion storage ring TSR of the Max-Planck
Institute for Nuclear Physics, which can store molecular ions of MeV energies
and is equipped with an electron cooler as well as an electron target employing
a cryogenic photocathode, so that electron-ion collision energies can be freely
set in the meV to eV range with unprecedented
resolution~\cite{electron-target-TSR}. While the main aim of the setup is to
provide a universal tool for measuring DR branching ratios of polyatomic
molecular ions into the different final fragment channels, the position
resolution is sufficient to allow also for 2D imaging of the breakup geometries.
The new setup has thus several advantages as compared to previous
approaches~\cite{PhysRevA1995-Datz-H2D+}, where grids of different transmissions
(yielding a set of linear combinations of the rates for individual fragment
channels) were employed in order to determine the branching ratios, a technique
which is usually only applied at zero relative electron-ion energies for
background reasons and which does not support molecular imaging.

The present paper discusses the concept and realization of the EMU detection
system and exemplifies its properties and possibilities by using the new
setup to investigate the
dissociative recombination of D$_2$H$^+$. The DR of \DDHi\ was studied before in
several storage ring experiments, and rate
coefficients~\cite{PhysRevLett2003-Lammich-D2H+,PhysRevA2008-Zhaunerchyk-D2H+},
the fragmentation geometry of the three-body
channel~\cite{PhysRevA2004-Strasser-D2H+}, and
branching ratios~\cite{PhysRevA2008-Zhaunerchyk-D2H+} have been reprted. The
kinetic energy $Q_0$ released in the three fragmentation channels, for ions and fragments
in their ground states and relative ion-electron energies $E_e=0$~eV, are
\begin{equation}\label{eqn-reaction}
\mbox{D} _2 \mbox{H} ^+ + \mbox{e}^- \longrightarrow \left\{
\begin{array}{llr}
\mbox{H + D + D}   & (+4.67\,\mbox{eV}) & (\alpha)\\
\mbox{HD + D}      & (+9.19\,\mbox{eV}) & (\beta )\\
\mbox{D} _2 + \mbox{H}     & (+9.23\,\mbox{eV}) & (\gamma)\\
\end{array}\right. .
\end{equation}
The branching ratios measured at $E_e \sim 0$~eV were reported to be
$B_\alpha$=76.5(2.2)\,\%, $B_\beta$=13.5(1.5)\,\%, and
$B_\gamma$=10.0(0.7)\,\%~\cite{PhysRevA2008-Zhaunerchyk-D2H+}.

\section{\label{basic-concept} The detector concept}

Since the beam energies $E_B$ of the molecular ions stored in the TSR are very
large as compared to the release energies occurring in DR experiments, the
fragments resulting from a DR event of a molecular ion of mass number $A$ are
traveling within a narrow cone around the ion beam axis, and the kinetic energy
$E_i$ of the fragment $i$ is to a very good approximation given by
\begin{equation}
E_{i} = \frac{A_i}{A} E_B 
\end{equation}
and is thus a measure of the mass number $A_i$ of the fragment. By recording separately the
energies of all DR fragments of an DR event with a large area multi-hit-capable
detector the fragmentation channel can be uniquely identified.

\begin{figure}[b]
  \includegraphics[width=8cm]{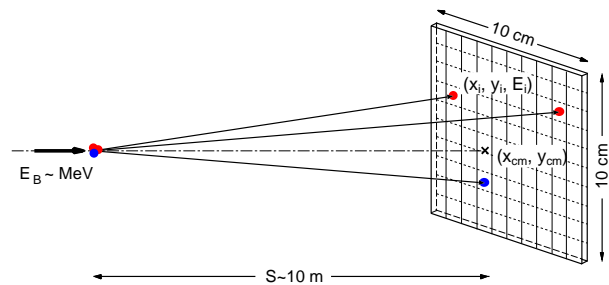}
  \caption{\label{fig-concept} Concept of the EMU detector: 
    A fast, neutralized molecule is dissociating at a distance $S$
    from a position- and energy-sensitive detector. The fragments
    continue to travel approximately at the velocity of the molecule,
    but due to the kinetic energy released in the dissociation they 
    acquire macroscopic distances from the transversal center-of-mass
    ($x_{cm}$,$y_{cm}$) when hitting the detector. The position
    ($x_i$,$y_i$) and the energy $E_i$ of each fragment is recorded.}
\end{figure}

In the present setup the energy-sensitive multi-hit detector is realized by
a large area (10$\times$10\,cm$^2$) position-sensitive, double-sided Si-strip
detector (see Fig.~\ref{fig-concept}). The energy- and position-sensitivity to
particles hitting the detector is achieved through 128 vertical (x-) and 128
horizontal (y-) strips on the front and back side of the detector,
respectively, which are read out individually by preserving the pulse height,
i.e., the energy information. The readout system also ensures via the timing
information that the detected particles belong to a single DR event. The
detector is mounted about $S$~$\approx$~10\,m downstream of the electron target
outside of the magnetic lattice of the TSR such that it can only be hit by
neutral fragments.

The impact of $n$ fragments from a single DR event on the detector leads to
$n_x$ responding x-strips $x_i$ and $n_y$ responding y-strips $y_i$ with
corresponding energies $E_{x_i}$ and $E_{y_i}$, and the detection of all
fragments can be easily verified by checking whether the two energy relations  
\begin{equation}
\label{energies}
\sum\limits_{i=1}^{n_x} E_{x_i} = E_B  \mbox{  and  }  \sum\limits_{i=1}^{n_y}
E_{y_i} = E_B 
\end{equation}
are fulfilled. 
The measured fragment mass numbers $A_{x_i}$ and $A_{y_i}$ corresponding  to
the energies recorded on the individual strips
are given by
\begin{equation}
\label{masses}
A_{x_i} = A E_{x_i} / E_B \mbox{  and  }
A_{y_i} = A E_{y_i} / E_B.
\end{equation}
In the following we will refer to the set of measured mass numbers  $A_{x_i}$,
$A_{y_i}$ and the corresponding strip coordinates $x_i$, $y_i$ as the 'hit
pattern' of the DR event.

The coordinates of the center-of-mass $(x_{cm}, y_{cm})$ of the detected event
in the detector plane are given by
\begin{equation}
\label{cm-coordinates}
(x_{cm}, y_{cm}) = 1/A \left(\sum\limits_{i=1}^{n_x} A_{x_i} x_i,
\sum\limits_{i=1}^{n_y}
A_{y_i} y_i \right) ,
\end{equation}
and an estimate for the transversal kinetic energy release $E_\perp$ can be 
determined from the projected weighted distances $D^2$ defined by
\begin{equation}
\label{distance}
D^2 = \sum\limits_{i=1}^{n_x} \frac{A_{x_i}}{A} (x_i-x_{cm})^2 +
\sum\limits_{i=1}^{n_y} \frac{A_{y_i}}{A} (y_i-y_{cm})^2
\end{equation}
using
\begin{equation}
\label{transversal-energies}
E_\perp  =  E_B \times D^2 / S^2,
\end{equation}
where $S$ is the distance between the center of the electron target and the detector.

While Eqs.~(\ref{energies})-(\ref{transversal-energies}) are valid even in cases
where more than one fragment hit the same strip, the strip-wise readout of
only two coordinate planes together with the finite strip width leads to
ambiguities in the determination of the breakup geometry from the hit pattern,
which in some cases also affects the identification of the fragmentation
channel. We shall first discuss these ambiguities for the specific case of the
DR of D$_2$H$^+$, before we consider the more general case.

\begin{figure}
  \includegraphics[height=6cm]{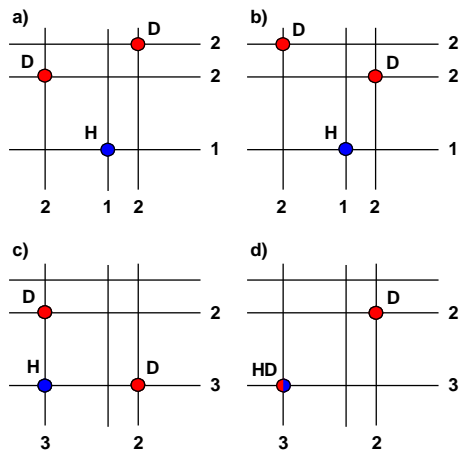}
  \caption{\label{fig-ambiguity} Some breakup geometries and corresponding hit
pattern occurring in the DR of D$_2$H$^+$. The lines mark the responding
read-out strips in vertical ($x_i$) and horizontal ($y_i$) directions; the
numbers at the end are the mass numbers $A_{x_i}$ and $A_{y_i}$
derived with Eq.~(\ref{masses}). The breakup geometries shown in a) and c)
result in the same read-out pattern (hit pattern) as the geometries shown in b)
and d), respectively. But while the hit pattern resulting from the
geometries shown in a) and b) still allow unique identification of the fragmentation
channel, the pattern resulting from the breakup geometries depicted in
c) and d) is ambiguous with respect to the fragmentation channels D+D+H and
HD+D.}
\end{figure}

Figure~\ref{fig-ambiguity} depicts the two most frequent of these ambiguities.
The hit pattern shown in
Fig.~\ref{fig-ambiguity}(a) is indistinguishable from the hit pattern shown in
Fig.~\ref{fig-ambiguity}(b); this pattern thus results in two solutions concerning
the fragmentation geometry. However, the identification of the fragmentation
channel is still unique. This is no longer the case for the hit pattern shown in
Fig.~\ref{fig-ambiguity}(c), where the H and the two Ds from the 3-body
fragmentation channel are forming an L-shaped pattern with the H in the corner such that only two
horizontal and two vertical strips are responding.
This pattern cannot be distinguished from that shown in
Fig.~\ref{fig-ambiguity}(d), which results from the detection of a molecular HD-
and an atomic D-fragment from the 2-body channel ($\beta$). Besides the even less
likely occurrence of two or more fragments hitting the same x- {\it and}
y-strip, i.e., the same ($x,y$)-pixel, these L-shaped patterns are the only
patterns in the DR of D$_2$H$^+$, which cannot be attributed uniquely to a
fragmentation channel. Since the probability for the occurrence of
this pattern is small due to the narrow strip width of $730$~$\mu$m
realized in the present setup, we attribute these patterns to the
channel with the smaller number of fragments, that is for the case discussed in
Figs.~\ref{fig-ambiguity}(c,d) to the HD+D channel. The influence of these
miss-assignments on the accuracy of the branching ratios is small and can
moreover be corrected for with the help of Monte Carlo simulations (see also
Sec.~\ref{sec-br-0eV}). 

While the ambiguities caused by fragments hitting the same ($x,y$) pixel discussed above are
present in case of all molecules, the other ambiguities depend on the multiplicity 
of identical fragment masses in the open fragmentation channels. In the DR of 
polyatomic molecules consisting of atoms of different masses such as HCO$^+$, 
the identification of the fragmentation channel and of the fragmentation 
geometry is not subject to any other ambiguities. In the DR of polyatomic 
molecules, which contain several atoms of the same mass and fragment into 
channels containing two or more identical fragments, the main ambiguities 
hampering the identification of the fragmentation channels are again
due to the L-shaped (sub-)pattern involving two identical
fragments at the two ends of the L. The interpretation of hit patterns in
terms of the fragmentation geometries is subject to similar ambiguities as
discussed above, but the number of geometries leading to a specific
pattern will strongly increase with the number of identical fragments.
However, we again would like to point out that the center-of-mass determination
by Eq.~(\ref{cm-coordinates}) and the transverse energy
distribution defined by Eq.~(\ref{transversal-energies}) are independent of
these ambiguities.

\section{\label{experiment-setup} Experimental setup}

\subsection{\label{EMU} The EMU detection system}

\begin{figure}
  \includegraphics[height=3.0cm]{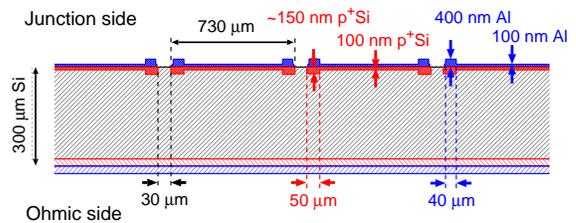}
  \caption{\label{fig-detector-cross} Schematic cross section through the EMU
detector, a double-sided, fully depleted Si-strip detector of $10 \times
10$\,cm$^2$ active surface area. The p$^+$Si and the Al
layers of the junction contact, through which the particles
enter the detector, were made as thin as possible to minimize the energy
losses of the DR fragments in these dead layers.}
\end{figure}

The large area double-sided Si strip detector employed in the EMU detection
system was built by the UK physics company Micron Semiconductor Limited~\cite{micron}. The detector has an
active area of 97.3\,mm$\times$ 97.3\,mm and an active depth of 300~$\mu$m when
biased by $\le -70$~V (Fig.~\ref{fig-detector-cross}). Position sensitivity has
been achieved by electronically subdividing the junction side into 128
vertical strips of 730~$\mu$m width, which are separated by gaps of 30~$\mu$m,
and the ohmic side in 128 horizontal strips of 700\,$\mu$m width, which are separated by
gaps of 60\,$\mu$m.

The thickness of the entrance (junction) window, which consists of a 
100\,nm-thick p$^+$ doped Si layer covered with a 100\,nm-thick aluminum 
coating, was minimized in order to keep the energy loss in the inert window 
material, and thus the cut-off energy for the detection of the heavy DR fragments, 
as small as possible. The increased thicknesses of the p$^+$Si and the Al layer 
at the edges of the strips, which ensure a clean separation between the strips 
and a loss-free readout and which affect about 14~\% of the active area of the 
detector, result in additional, slightly down-shifted peaks in the
pulse height spectra; particles traversing this part of the entrance window
suffer a larger energy loss which is specific for the fragment energy, mass, and
its nuclear charge. This is discussed in more detail in Sec.~\ref{Data processing}
in connection with the pulse height spectrum (Fig.~\ref{fig-adc}) observed in
the DR of D$_2$H$^+$. Since the penetration depth of the fragments
in the detector material for energies
available at the TSR is in most cases less than 10~$\mu$m and
the gap region close to the surface is almost field-free, fragments impinging in
the gap region, covering about 4~\% of the detector surface, will likely not be
detected.

\begin{figure}
  \includegraphics[width=8cm]{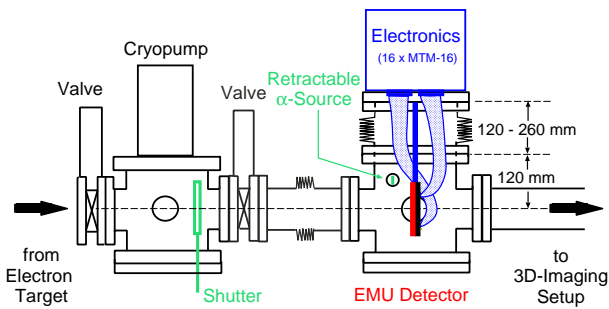}
  \caption{\label{chamber} Schematic view of the EMU setup, which is located in
the BAMBI beamline downstream of the electron target of the TSR. }
\end{figure}

The detector is mounted in a dedicated chamber (Fig.~\ref{chamber}) in the
neutral fragment beamline BAMBI downstream of the electron
target~\cite{electron-target-TSR} of the TSR. The distance from the center
of the electron target to the front face of the detector is $S_0=941(1)$~cm. At
this distance the detector covers the full cone size for the DR fragments that
is allowed by the beamtube (CF100). The maximum visible cone is in fact limited
vertically by the height of the vacuum chamber in the TSR dipole magnet
to about 8\,cm diameter. A large bellow allows to retract the detector out of
the beamline so that experiments with detectors further down the beamline are
possible without breaking the vacuum. Moreover, the detector can be moved up and
down in front of a retractable $\alpha$-source, which in turn can be moved
horizontally using a translational stage, so that any part of the detector can
be irradiated for testing and calibration purposes. The cryopump is
vibrationally decoupled from the detector chamber to protect the wire bonds to
the 256 strips of the detector from the mechanical noise of the pump. Under
running conditions the vacuum in the detection chamber is
about $5 \times 10^{-9}$~mbar. A mechanical shutter is furthermore
installed in the pumping chamber to protect the detector during the injection
phase of the ion beam into the storage ring.
\begin{figure} [b]
  \includegraphics[width=8.5cm]{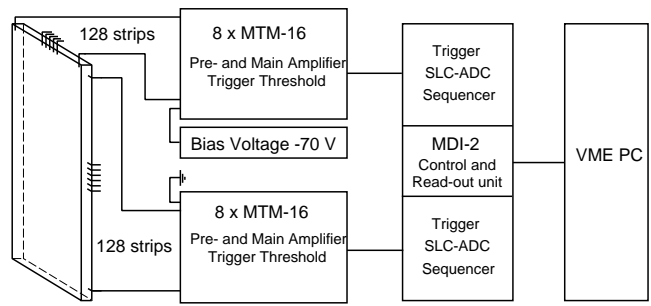}
  \caption{\label{fig-electronics} A schematic view of the electronic setup for
the EMU detector.}
\end{figure}
The 256 strips of the detector are read out by 16 highly integrated MTM-16
units developed by Mesytec~\cite{mesytec}, which are mounted directly on the
upper lid of the detector chamber to minimize the length of the cables between
detector and preamplifier. Each unit contains 16 channels consisting of a
preamplifier, main amplifier, shaper and hold stage. Moreover, four input signals
are added and compared to an adjustable discriminator level. The eight boards
connected to the 128 x-strips and the eight boards serving the
y-strips are daisy-chained to the two function blocks, respectively, of an MDI-2
unit developed by Mesytec as well. This one-slot VME unit allows one to do all
the controlling, timing and read out of the 16 MTM-16 boards. A trigger
signal is produced when one of the discriminators of the 16 MTM-16 units
responded; it activates the hold stages of the MTM-16 units after ~2~$\mu$s,
(i.e., only signals arriving within this time will be held) and starts the
readout sequence. The individual pulse heights are digitized by a sliding scale
ADC and - if above a programmable threshold - stored together with the strip
number in a FIFO, which is read out via the VME interface. In the present
configuration the readout speed is limited to about 2000 events per second.
A schematic view of the electronic set up is shown in
Fig.~\ref{fig-electronics}.

\subsection{\label{DDHi} Specific settings for measuring the DR of \DDHi}

The first application of the EMU detection system was the investigation of the 
dissociative recombination of \DDHi. \DDHi\ ions were produced in a Penning source,
accelerated to $E_B = 2.39$\,MeV using an rf-quadrupole accelerator, and
injected into the TSR; the stored ion current was up to about 4\,$\mu$A. After
injection the ion beam was first phase-space cooled by the velocity-matched
electron beam supplied by the electron target, which had a density of
9$\times$10$^6$\,cm$^{-3}$ and a transversal temperature of approximately
1.5\,meV (expansion factor 20). At these conditions the effective length of the
electron target is 115\,cm, with an additional 20\,cm on either end where the
electron beam is merged with and demerged from the ion beam. After a few seconds
the ion beam was phase-space cooled to a beam diameter below 1.0\,mm.

The electron beam was then used as a target, either at matched or detuned
electron velocities. The detuning energy $E_d$, which measures the average
electron-ion collision energy in the center-of-mass system, was adjusted  to 
values between 0 and 20\,eV by changing the acceleration voltage of the 
electron beam. At detuning energies $E_d > 0$ the electron beam velocity was 
switched at a rate of 10 - 100~Hz between $E_d$ = 0 and the required energy 
$E_d$ in order to keep the ion beam phase-spaced cooled. Data was recorded 
between a few seconds and up to 30\,s after injection, and count rates were 
limited to $\le 2$\,kHz.

\section{Data analysis}
\label{Data processing}

The first analysis step consists of shifting the pulse height spectra of the
individual strips to a common scale in order to compensate for the slightly
different amplifications and offsets of the individual electronic channels using the \DDHi\ data. 

\begin{figure}
\includegraphics[width=8cm]{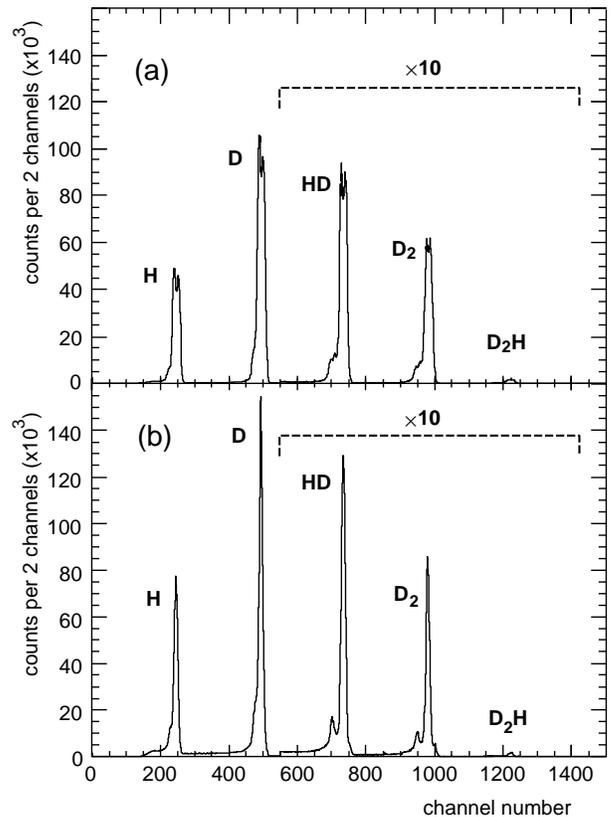}
\caption{\label{fig-adc} Pulse height spectra obtained in the DR of 2.39 MeV \DDHi\ ions
at a detuning energy of
$E_d = 0$~eV. The figure shows the sum of the calibrated pulse height
spectra (a) of all x-strips on the front and (b) of all y-strips on the back
side. Above channel 550, counts are multiplied by a factor of 10.}
\end{figure}

The sums of all 128 x- and of all 128 y-strip spectra performed after
this calibration step, are displayed in Fig.~\ref{fig-adc}. While
the overall energy resolution reached with the y-strips on the back side of the
detector (e.g., $\Delta  E$(FWHM)$ < 30$~keV for Ds of 0.96~MeV) is already very
satisfactory, the resolution of the front side strips is worse by almost a
factor of two. This is mainly due to the excess noise observed on the
front side strips; while the back side of the detector is electronically 
well-shielded by a metal plate mounted directly behind the detector, the front 
side is unshielded in the direction of the storage
ring. But although the resolution of the x-strips is not yet perfect, the peaks
corresponding to the detection of an H, D, HD (or H+D), and D$_2$ (or D+D) can
be clearly identified. Even the small peak is visible that is caused by
the detection of all fragments (D$_2$H) on a single strip.

On the low-energy side of the peaks small satellite lines are visible. As
already pointed out in Sec.~\ref{EMU} these satellites are caused by particles
passing through the slightly thicker parts of the entrance window.
For light fragments as in the present
case of the DR of \DDHi\, the resulting energy difference between the main and the
satellite peak is small and thus does not hamper the mass assignment. For heavier
fragments like carbon or oxygen, the energy difference will be larger and can
correspond to one or more mass units. But even in these cases this will not lead
to a misinterpretation of a DR event as the two energy sums [see
Eq.~(\ref{energies})] will not be fulfilled. Disregarding these
events, however, reduces the detection efficiency of the set-up and has to be
taken into account by detailed simulations.

A more careful examination of Fig.~\ref{fig-adc}(b) reveals a small number of
counts between the peaks. These are mainly caused by events where the charges
created by a fragment in the active volume of the detector are drifting to two
adjacent y-strips such that the charge collected by each strip results in a
reduced signal height. Such a pulse height splitting can be at least partly
reconstructed by adding the signals of the two adjacent y-strips.
A corresponding pulse height splitting on the front side of the detector is 
very unlikely to occur [see also Fig.~\ref{fig-adc}(a)] as the range of
the fragments in the detector material is so small that fragments hitting the
gap between the x-strips are likely not be detected at all. As in the case
discussed above, the influence of these effects on the
detection efficiency is taken into account by simulations.

After the calibration procedure the pulse height of each of the 256 strips is
compared to appropriately chosen windows around the mass peaks and, if it falls
into one of these windows, the corresponding mass number is assigned to the
strip. If the pulse height is above the lower
threshold but does not fall into one of the mass windows, it is tested whether
the sum with one of the neighboring strips fulfills this criterion. If yes,
the corresponding mass is assigned to the strip with the larger signal
while the other strip is ignored in the further analysis. We shall refer to this
set of  $A_{x_i}$ and $A_{y_j}$ and adjusted positions $x_i$ and $y_j$, which are
randomized across the width of the respective strip to avoid binning problems, as a hit pattern.

The next analysis step consists of checking whether the sums over all mass numbers
$A_{x_i}$ and $A_{y_j}$, respectively, observed in an event are equal to the mass
number of the dissociating molecule, and whether the hit pattern corresponds to
a possible DR fragmentation channel, i.e., whether there is any combination of
the atomic constituents of the dissociating molecule that can explain the observed hit pattern.
If yes, the event is marked as a DR event and the final fragment channel is
assigned. Here, preference is given to the fragmentation channel with the
smallest number of fragments. In the case of the DR of \DDHi, for example, this
means that the rather unlikely L-shaped hit pattern shown in
Fig.~\ref{fig-ambiguity}(c), which results from the three-body breakup and
cannot be distinguished from the hit pattern caused by the two-fragment channel
HD+D as the comparison with Fig.~\ref{fig-ambiguity}(d) shows, is assigned to
the HD+D channel. 

\begin{figure}
\includegraphics[width=8cm]{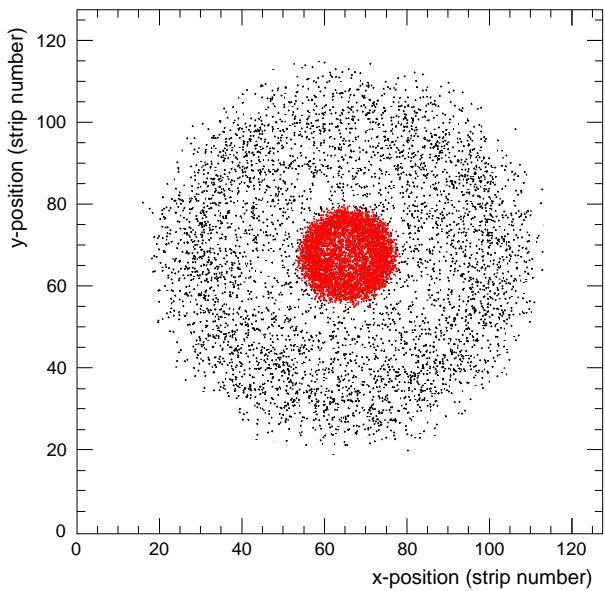}
\caption{\label{fig-hydrogen} The distribution of the H (black) and D$_2$ (red)
positions observed for 5000 $D_2$+H events following the DR of
\DDHi\ at $E_d$=0\,eV. Note that H events within the red area are partly
shadowed by the D$_2$ events due to the plotting procedure. }
\end{figure}

An example for the extracted position distribution of fragments for a single
fragment channel is given in Fig.~\ref{fig-hydrogen}, where the positions of the
D$_2$ (red) and the H (black) fragment of the D$_2$+H channel following the DR
of \DDHi\ at $E_d = 0$~eV are shown in form of a scatter plot. One can clearly
see the two circular shapes corresponding to the fragments with
mass numbers 1 and 4, respectively.

\begin{figure}
\includegraphics[width=8.5cm]{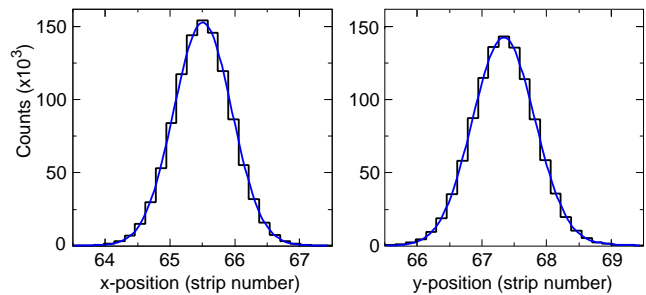}
\caption{\label{fig-cm} The horizontal and vertical distribution of the
center-of-mass coordinates of DR events of \DDHi\ taken after 30\,s of
phase-space cooling at collision energies of $E_d$ = 0\,eV. The blue
lines are fits by a Gaussian distribution.}
\end{figure}

From the positions  $x_i$ and $y_j$ and the mass numbers $A_{x_i}$ and $A_{y_j}$
the (transversal) center-of-mass position ($x_{cm}$, $y_{cm}$) is calculated.
The resulting x- and y-distributions of the center-of-mass position for events
following the DR of \DDHi\ at $E_d$ = 0\,eV are shown in Fig.~\ref{fig-cm}.
The data was taken 30\,s after injection and continuous electron
cooling, and exemplifies the good phase-space cooling that can be reached  with
the electron target of the TSR; taking the width of the strips into account the
center-of-mass spread at the position of the detector is estimated to be 
$\lesssim 0.9$~mm (FWHM) in x- as well as in y-direction. In the further analysis 
of the data only events within a 2$\sigma$-ellipse around the centroid of the 
center-of-mass distribution are used to further suppress background events which 
might result from dissociative charge exchange reactions of the molecular ion 
with the residual gas of the TSR.

To investigate the influence of the detector properties and of the
approximations and cuts used in the data analysis on the deduced results, and
to allow for a detailed comparison with theoretical predictions
for the DR process, a Monte Carlo simulation program has been developed. The simulation
creates DR events at random position in the overlap region between the molecular
ion beam and the electron beam and propagates them towards the detector.
Subroutines are available to allow for an internal excitation of the ion and to
incorporate different kinetic energy releases and angular distributions of the
fragments with respect to the center-of-mass direction. At the detector, the
impact position for each fragment is determined, which includes the decision
as to which part of the entrance window was hit. The
kinetic energy loss in the appropriate dead layer is
subtracted for each fragment and a pulse proportional to the remaining kinetic
energy is created on the strip being hit, broadening the pulse height by a
Gaussian shaped resolution function. For fragments hitting the
same strip the pulse heights are added. If the gap between two
x-strips is hit, the fragment is assumed not to be detected. If
the impact position happens to be in the gap between two y-strips of the back
side, the signal is shared. The output of the program for a simulated event is
a list of strip numbers and corresponding pulse heights. The simulated
data are then treated with the same analysis tools as the measured data.

\section{Results for the DR of \DDHi}

\subsection{\label{sec-br-0eV} Branching ratio at $E_d=0$~eV}

The branching ratios for the three final fragment channels accessible in the DR
of \DDHi\ at relative ion-electron energies of $E_d=0$~eV were determined by
analyzing data taken after 30-50\,s of electron cooling. Although the phase space
cooling of the ion beam and the vibrational cooling of the ions is achieved already
after $\sim 3$~s of electron cooling, the cooling was continued in order to lower the
rotational temperature of the ions to values below the ambient room temperature of
300~K as shown in Ref.~\cite{PhysRevLett2003-Lammich-D2H+}. The
relative number of DR events leading to the three fragmentation channels (see
Eq.~(\ref{eqn-reaction})) are $N_\alpha$=76.69(4)\,\%, $N_\beta$=14.32(4)\,\%,
and $N_\gamma$=8.99(3)\,\%, where the uncertainties given are of purely
statistical origin. The Monte Carlo simulation was used to correct the measured
numbers for the slightly different efficiencies for detecting and identifying
the fragment channels. In this simulation the angular distributions of the
fragments were assumed to be isotropic, but the population of the different
vibrational states of the HD and D$_2$ fragments as determined in
Sec.~\ref{two-body} was explicitly taken into account. The resulting
corrections which had to be applied to the measured relative numbers given
above amount to less than two percentage points.

The deduced branching ratios $B$ are compiled in Table~\ref{tab-br}. The
uncertainties given are dominated by systematic errors caused mainly by the not
yet fully understood detection efficiency for fragments hitting the gap regions.
The ratios are compared to the result of a recent measurement performed at
CRYRING~\cite{PhysRevA2008-Zhaunerchyk-D2H+} using the grid method. Within the
error bars the two measurements agree well despite the presumably different
initial rotational temperatures of the \DDHi\ ions (see also
the following section). As discussed already in
Ref.~\cite{PhysRevA2008-Zhaunerchyk-D2H+} the branching ratios display a
clear isotope effect: On a purely statistical ground the HD+D channel
should be twice as strong as the D$_2$+H channel, i.e., in the
absence of an isotope effect one would expect 2$B$(D$_2$+H)/$B$(HD+D)$ = 1$.
However, the measured values result in a deuteration enhancement of 2$B$(D$_2$+H)/$B$(HD+D)$ =
1.27(0.05)$, slightly smaller than deduced in
Ref.~\cite{PhysRevA2008-Zhaunerchyk-D2H+} but very similar to the ratio
$B({\mbox{HD+H}})/2B({\mbox{H}_2+\mbox{D})} = 1.20(0.05)$ observed in the DR of
H$_2$D$^+$~\cite{PhysRevA1995-Datz-H2D+}. These results clearly show that in the
DR of deuterated H$_3^+$ the formation of the two-body fragment channels
containing the most deuterons is enhanced.

\begin{table}
\caption{\label{tab-br} Branching ratios obtained in the present experiment
for the three fragment channels following the DR of \DDHi\ at $E_d$ = 0\,eV. The
results are compared to those of a recent measurement at
CRYRING~\cite{PhysRevA2008-Zhaunerchyk-D2H+}.
 }
\begin{ruledtabular}
\begin{tabular}{lcc}
Channel & This work & CRYRING\cite{PhysRevA2008-Zhaunerchyk-D2H+}\\
\hline  
($\alpha$) H+D+D   & 78.0(0.4) & 76.5(2.2)\\
($\beta$)  HD+D    & 13.5(0.3) & 13.5(1.5)\\
($\gamma$) D$_2$+H &  8.6(0.2) & 10.0(0.7)\\
\end{tabular}
\end{ruledtabular}
\end{table}

\subsection{\label{sec-br} Branching ratios at $E_d > 0$~eV}

To determine the branching ratios into the three fragment channels at relative
ion-electron energies  $E_d > 0$~eV, data collection was started after 3 seconds
of electron cooling when the ion beam was already fully phase space cooled.
Moreover, at this time all ions are vibrationally relaxed while the rotational
temperature is expected to be around 300~K or even slightly 
higher~\cite{PhysRevLett2003-Lammich-D2H+}. Comparing the
branching ratios at $E_d=0$\,eV obtained after the shorter phase space cooling 
with those discussed in Sec.~\ref{sec-br-0eV}, no difference was noticable 
within the limits of precision.
After electron cooling, the energy of the electron beam was frequently
switched between the cooling energy and the desired detuning energy $E_d > 0$~eV, changing the
latter from injection to injection to span the range between 0 up to 20 eV. The
detuning-energy cycle was repeated several times to collect statistically 
reliable data. The branching ratio at each detuning energy
was then obtained as described in Sec.~\ref{sec-br-0eV}, including the corrections
due to the slightly different detection efficiencies for the three fragment
channels. Efficiency losses that could occur at high
detuning energies due to the finite size of the detector
were shown to be insignificant by investigating the distribution of impact positions of
the fragments on the detector; at high detuning energies a considerable part of the
maximal available release energy is not converted into
kinetic energy of the fragments, but into internal excitation energy of the fragments.
Moreover, background events from dissociative charge exchange collisions with
the residual gas were found to be negligible at all measured ion-electron
energies.

\begin{figure}
\includegraphics[width=8cm]{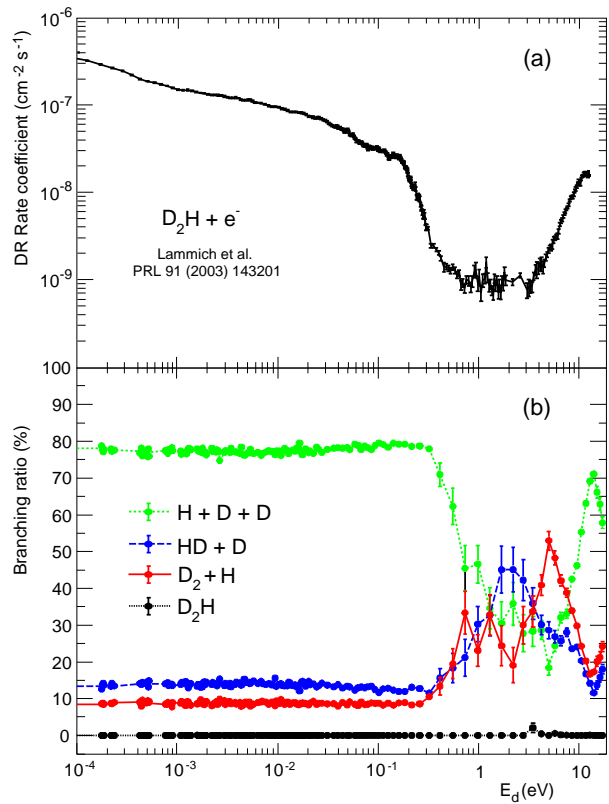}
\caption{\label{fig-br} (a) Energy dependence of the DR rate coefficient of
\DDHi\, as measured by Lammich {\it et al.}~\cite{PhysRevLett2003-Lammich-D2H+}.
(b) Fragment channel branching ratios determined in the present work as a
function of the collision energy with only statistical uncertainties. 
In the absence of isotope effects one expects to find $B$(HD+D) = 2$B$(D$_2$+H).
}
\end{figure}

Figure~\ref{fig-br} shows the extracted fragment channel branching ratios in the
range of $E_d$ = 0.2\,meV to 20 eV together with the DR rate coefficient
measured in a previous experiment at the TSR under similar cooling
conditions~\cite{PhysRevLett2003-Lammich-D2H+}. As to be expected, the
probability for the recombination channel D$_2$H, which
requires the emission of a stabilizing photon in times comparable to the
dissociation times, is found to be consistent with zero to within better than
1\%. The branching ratios into the fragmentation channels are observed to
be more or less constant up to $E_d\approx$ 300\,meV at values close to those
obtained at $E_d=0$\,eV, even though the DR rate coefficient drops by two orders of
magnitude.  In particular, the isotope effect observed at $E_d=0$~eV
persists over this energy range and is even increasing for energies $\gtrsim
300$~meV. For higher energies the branching into the three-body channel is
declining quickly and reaches a minimum level of about 20\,\% between 4 and 5\,eV, while the relative intensities of the two-body channels are both
rising. At $E_d$=2\,eV the HD+D channel is the dominant fragmentation
channel  ($B$(HD+D)$ \sim $45\,\%), but at around 5~eV the D$_2$+H channel is
even reaching $B$(D$_2$+H) $\sim$ 55~\%, exhibiting a huge isotope effect
of 2$B$(D$_2$+H)/$B$(HD+D)$ = 3.7(0.5)$. At energies above 10~eV the
three-body channel regains its role as the dominant fragment channel with
a branching ratio above 60\,\%.

We are not aware of earlier measurements of the fragment branching ratios
following the DR of \DDHi\ at detuning energies $E_d > 0$~eV. However, Datz {\it
et al.}~\cite{PhysRevLett1995-Datz-H3+,PhysRevA1995-Datz-H2D+} have measured
these ratios in the energy range between $\sim$1\,meV and 20\,eV for the DR of
H$_3^+$ and its isotopomer H$_2$D$^+$. The overall behavior of the relative
branching ratio between the combined two-body channel and the three-body channel
is rather similar for all three systems investigated, and the gross
structure of this ratio has been successfully attributed to the successive
opening of the electronically excited two-body fragment channel
H($nl$)+H$_2(X^1\Sigma_g)$, H($2l$)+H$_2(B^1\Sigma_u)$ and of the three-body
channel H($2l$)+H($1s$)+H($1s$) within a statistical
approach~\cite{PhysRevA2001-Strasser-H3+vib}. However, while in the DR of
H$_2$D$^+$ the relative fractions of the two two-body channels seem to stay
constant over the measured range of energies, we find that this fraction 
strongly changes in the DR of \DDHi\ at energies between $\sim 1 - 10$~eV,
leading to a drastic isotope effect around $E_d \sim 5$~eV that is yet to be explained.

\subsection{\label{two-body} Vibrational excitation of fragment molecules}

Standard imaging techniques without mass identification have difficulties
identifying and separating two-body fragmentation channels in DR experiments
involving polyatomic molecular ions. The event-by-event mass identification of
all fragments provided by the EMU system allows for a straight-forward analysis of
these channels.

\begin{figure}
\includegraphics[width=8.cm]{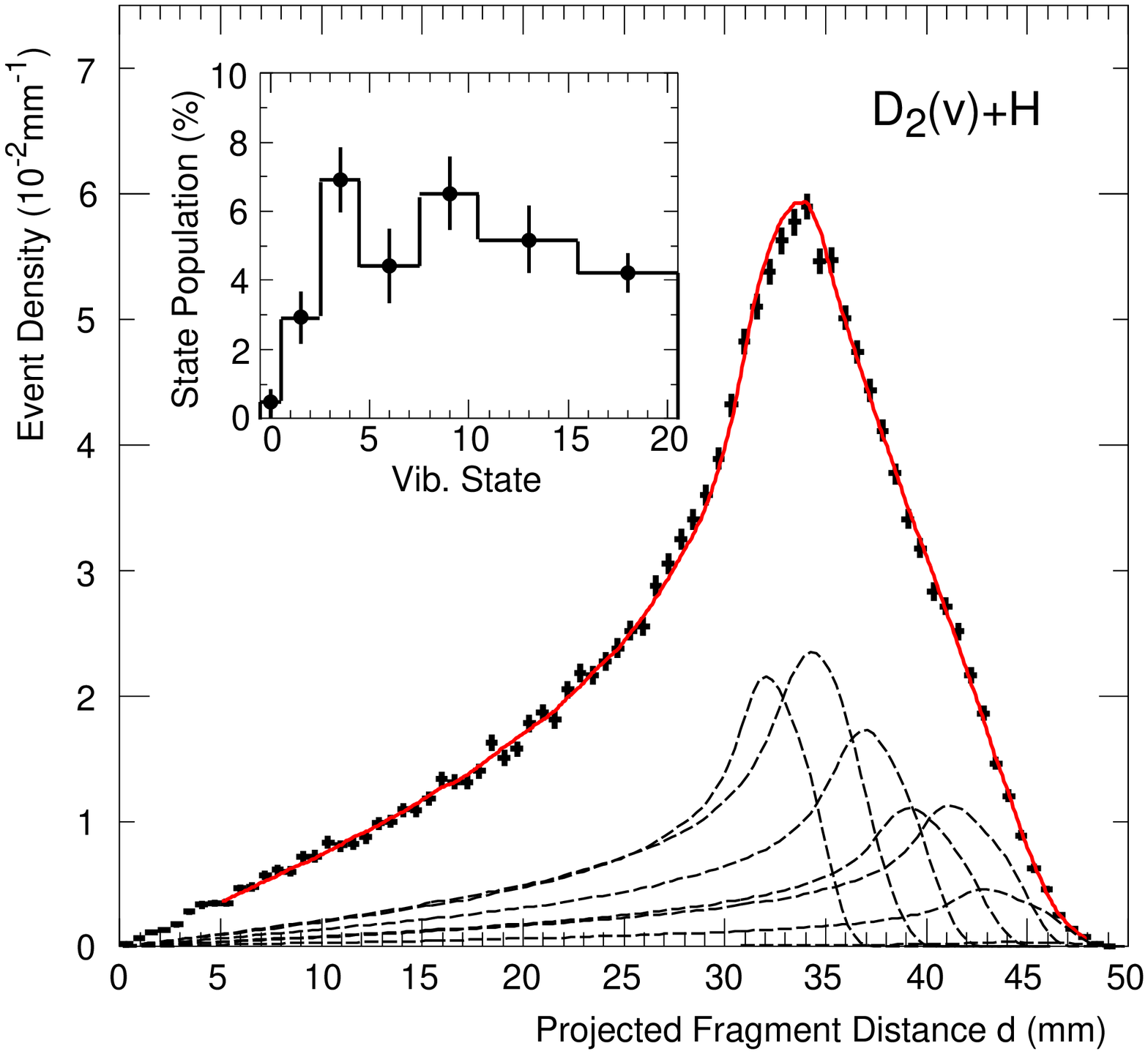}
\includegraphics[width=7.8cm]{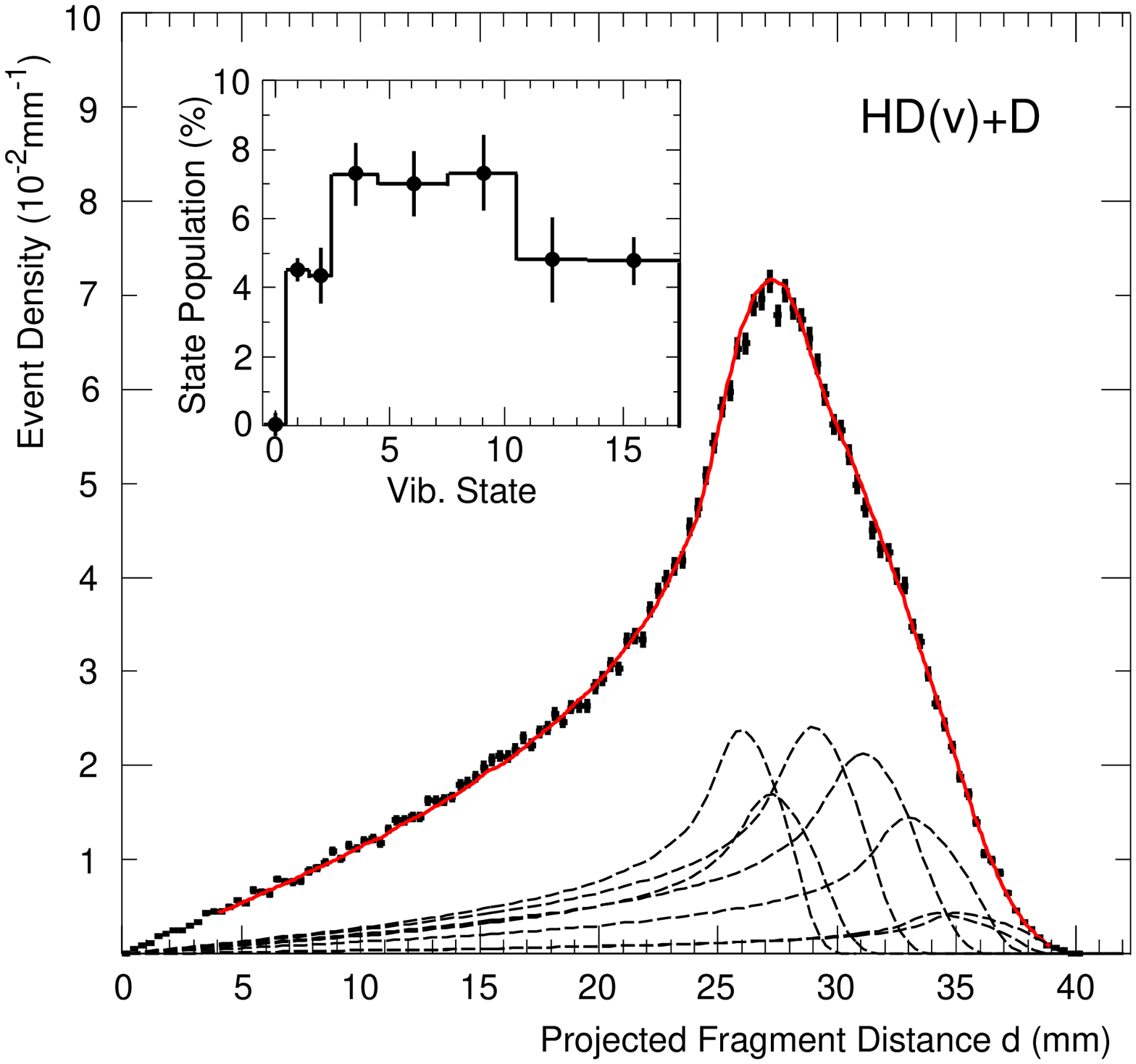}
\caption{\label{fig-2body} Normalized projected distance distributions for the
two-fragment channels D$_2$+H (upper panel) and HD+D (lower panel) observed in
the DR of \DDHi\ at $E_d=0$~eV;  the solid lines are the result of a fit with
simulated distributions (dashed lines) corresponding to the population of
different vibrational levels  D$_2(v)$ and HD($v$), respectively. The resulting
population probabilities are shown in the inserts; for histogram bins spanning
more than one unit in $v$, the relative populations were forced to be equal
within the bin.}
\end{figure}

The normalized projected distance distributions $P(d)$ for the two two-body
fragment channels D$_2$+H and HD+D observed in the DR of \DDHi\ at $E_d=0$~eV
are displayed in Fig.~\ref{fig-2body} using data collected after
30~s of electron cooling. Following common practice (see e.g.,
Ref.~\cite{PhysRevA2002-Strasser-H3+D3+}), the distance
$d=((x_1-x_2)^2+(y_1-y_2)^2)^{1/2}$ between the two fragments in the detector
plane is plotted rather than the weighted projected distance $D$ defined by
Eq.~(\ref{distance}). The projected distance distributions $P(d)$ carry
information about the kinetic energy release as well as about the angular
distribution of the fragments with respect to the ion (electron) beam direction~\cite{novotny}. Since at
$E_d=0$~eV the angular correlations are expected to be isotropic for symmetry
reasons, $P(d)$ can be readily analyzed to yield the kinetic energy release distribution.

In both channels only the vibrational states built on the electronic ground
state of the fragments are energetically accessible, i.e., the available final states are
D$_2(X^1\Sigma_g(v))$+H$($1s$)$ with $v$\,=\,0 to 20 and HD$(X^1\Sigma_g(v))$+D($1s$)
with $v$\,=\,0 to 17, respectively. The relative populations of these states, which are
derived from a fit of the projected distance distributions $P(d)$ using
simulated distributions for individual vibrational states, are shown in the
inserts. In the simulation we assume the rotational temperature of the parent 
ion to be 100~K (see also further below) and the DR rate coefficients to be constant for all rotational
angular momenta. For these fits some vibrational levels were
grouped together and an equal population was assumed within each group
to reduce the uncertainties caused by the anticorrelation of contributions
between energetically close levels; they are nevertheless still dominating
the accuracy of the deduced populations.

The derived population distributions of the vibrational levels for the molecular
fragment are similar for both channels and similar to the distributions measured by
Strasser {\it et al.}~\cite{PhysRevA2002-Strasser-H3+D3+} in the DR of H$_3^+$
and D$_3^+$. They all show a bell shape-like behavior, close to what is
expected from a simple phase space argument~\cite{PhysRevA2001-Strasser-H3+vib}.

\subsection{\label{three-body} The three-body channel H+D+D}

Also in studies of DR fragmentation channels which result in a total breakup
of the molecule into its atomic constituents, the EMU system has some advantages
as compared to standard imaging techniques based on micro-channel plates and an
optical readout, even though the latter usually exhibits a considerably better
position resolution (e.g., $< 100$~$\mu$m~\cite{novotny} as compared to $\sim$ 750\,$\mu$m
of the EMU detector). Besides the unambiguous fragment identification, the 
advantage is in particular the high data rate of up to 2000\,Hz that can be 
handled by the EMU system (to be compared to, e.g., $\sim 30$~Hz of the TSR 
optical imaging setup~\cite{novotny}).

In the DR of \DDHi\ at $E_d=0$~eV the three atomic products can only be formed in
their electronic ground states. The total kinetic energy release $E_k$ in
the D+D+H channel is thus uniquely determined by the ground state energy release 
$Q_0$ of \DDHi\ with respect to the asymptotic free neutral atomic products
plus the remaining excitation energy of the \DDHi\ molecule,
which we describe by a (rotational) temperature. As discussed in
Sec.~\ref{basic-concept}, the weighted projected distance $D^2$, which is not
affected by any of the ambiguities in the fragmentation geometry determination, 
is proportional to the transverse energy release. The distribution of
the weighted projected distances $D^2$ thus provides information on the
total kinetic energy release.

\begin{figure}
\includegraphics[width=8.5cm]{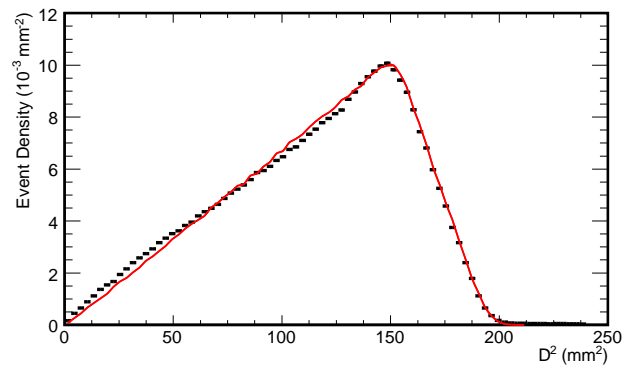}
\caption{\label{fig-ker-H+D+D} Normalized distribution of weighted transverse distances
$D^2$ observed in the DR of \DDHi\ at $E_d=0$~eV for the fragment channel H+D+D
after 30\,s of phase-space cooling. The solid line is the result of a simulation
assuming an isotropic breakup geometry and an initial rotational excitation of
the \DDHi\ ions corresponding to a rotational temperature of 100~K.}
\end{figure}

The measured distribution, derived from data collected after 30~s of electron
cooling, is plotted in Fig.~\ref{fig-ker-H+D+D} together with the simulated
distribution, which was calculated assuming that the three fragments are
isotropically distributed in the available phase space and also assuming
$Q_0=4.67$~eV and a rotational temperature of 100~K. While
the edge of the distribution at large $D^2$, which is most sensitive to the
total kinetic energy release, is very well described by the simulation
assuming a subthermal rotational temperature of 100~K (see also Ref.~\cite{PhysRevLett2003-Lammich-D2H+}), there
are small but statistically relevant deviations on the left shoulder of the
distribution, which are likely caused by anisotropies in the momentum distribution
between the three fragments. 

Although we only have access to the transverse kinetic energies of the
fragments, it has been shown
before~\cite{PhysRevA2002-Strasser-H3+D3+,PhysRevA2004-Strasser-D2H+} that
information about the momentum distribution between the three fragments can be
gained employing a slightly modified concept of Dalitz coordinates and
Dalitz plots. The coordinates $\eta_1$ and $\eta_2$
originally introduced by Dalitz~\cite{PhilosMag1953-Dalitz-Dalitzplots} are
linear combinations of the kinetic energies $E_i$ of the three fragments in the
center-of-mass frame, taking into account energy and momentum conservation. By
plotting the number of events as a function of $\eta_1$ and $\eta_2$, the
momentum correlations between the fragments can be visualized. In particular,
for a purely statistical, phase-space dominated breakup the Dalitz plot is
evenly populated. The modified
concept~\cite{PhysRevA2002-Strasser-H3+D3+,PhysRevA2004-Strasser-D2H+,lammich,PhysRevA2007-Nevo-CH2+}
consists in defining projected Dalitz
coordinates $Q_1$ and $Q_2$ by replacing $E_i$ by the corresponding transverse
energy $E_{\perp,i}$ and by plotting the number of events as a function of $Q_1$
and $Q_2$ instead of $\eta_1$ and $\eta_2$. The measured distribution is then
divided by a simulated distribution assuming a phase-space dominated breakup
in order to account for possible detection efficiency variations and to regain a
uniformly populated Dalitz plot in case no momentum correlations between the
fragments exist. We will refer to these plots as Dalitz ratio plots.

In the case of the DR of \DDHi\ where two of the three fragments are identical,
the projected Dalitz coordinates $Q_1$ and $Q_2$ are conveniently defined
as~\cite{lammich}
\begin{eqnarray}
Q_1 & = &  \sqrt{\frac{A}{A_H}}\frac{E_{\perp,D_2} - E_{\perp,D_1}}{3 E_\perp} \\
Q_2 & = &  \frac{A}{3 A_D} \frac{E_{\perp,H}}{E_\perp} - \frac{1}{3}
\end{eqnarray}
with
\begin{equation}
\label{individual transversal energy}
 E_{\perp,i}=\frac{A_{i}}{A} [(x_i-x_{cm})^2 + (y_i-y_{cm})^2] \frac{E_B}{S^2}
\end{equation}
and $E_\perp = \sum E_{\perp,i}$. The allowed values of
$Q_1$ and $Q_2$ are confined by $Q_1^2 +
Q_2^2 < $1/9.

\begin{figure}
\includegraphics[width=8cm]{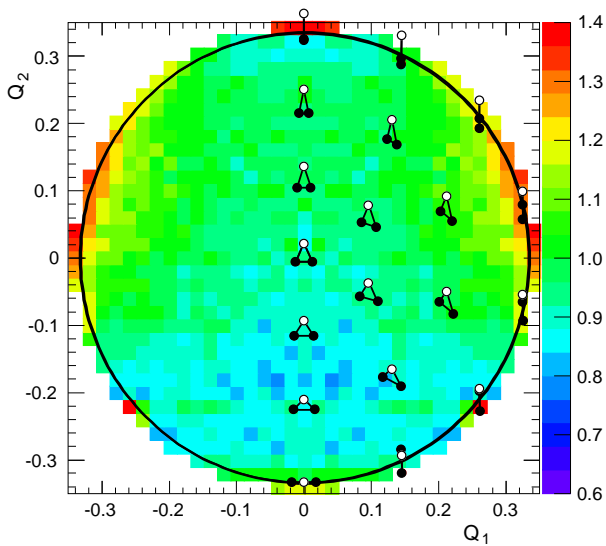}
\caption{\label{fig-dalitz} Dalitz ratio plot of the three-body
fragmentation channel following the DR of \DDHi\ at $E_d=0$ after 30~s of
electron cooling. The triangles depict projected dissociation geometries for a
sample of points in the $(Q_1,Q_2)$ plane; the white dot shows the position of 
the H, the black points the positions of the Ds. }
\end{figure}

In filling the Dalitz ratio plot two ambiguities have to be considered. The
first is connected with the indistinguishability of the two Ds, which leads to
two entries: at $(Q_1,Q_2)$ and $(-Q_1,Q_2)$. The second one is specific to the
EMU system and is caused by the ambiguity [see Figs.~\ref{fig-ambiguity}(a) and \ref{fig-ambiguity}(b) and
Eq.~(\ref{individual transversal energy})] in deducing the breakup geometry from
the hit pattern. This leads to two additional entries at $(Q'_1,Q_2)$ and
$(-Q'_1,Q_2)$ and to a smearing of structures along the $Q_1$ coordinate. The
resulting Dalitz ratio plot, divided by the distribution for an isotropic breakup
of D$_2$H$^+$, is shown in Fig.~\ref{fig-dalitz}. It is quite obvious that the
momentum distribution is far from being isotropic and that linear decay
geometries are enhanced by up to 40\%, where one of the Ds and the H are emitted back-to-back with the second
slow D remaining in the center.

The present result agrees with the result obtained in
Ref.~\cite{PhysRevA2004-Strasser-D2H+} using standard 2D-imaging, but
the almost two order of magnitude higher event rates that can be handled by the
EMU system are resulting in improved statistics and thus in a considerably
clearer picture. The statistics which can now be attained in these
measurements should finally be sufficient to employ the Monte Carlo image restoration
technique discussed in Ref.~\cite{PhysRevA2004-Strasser-D2H+} with strongly reduced artificial noise patches. This will allow for a
more quantitative analysis of the momentum distributions.

\section{Summary}

The potential of the EMU imaging system for studying the dissociative
recombination of polyatomic molecular ions with electrons in merged beam
experiments is clearly borne out by the results obtained for the DR of \DDHi. The
main advantage of the new set up, which is based on a large area, energy and
position sensitive Si detector with multi-hit capabilities, is the possibility
to determine the individual masses of the fragments on an event-by-event basis.
This allows one to efficiently distinguish DR events leading to only neutral
fragments from DR and background events involving charged fragments, to uniquely
identify these fragmentation channels, and to determine their branching ratios as
a function of the relative ion-electron energy. While the position resolution,
which is limited by the width of the read-out strips to $\sim 750$~$\mu$m,
cannot compete with the resolution of $< 100\,\mu$m of optical
imaging systems, it is nevertheless sufficient to allow for detailed studies of
total kinetic energy releases and transverse breakup geometries of the
different fragmentation channels by 2D imaging; the lack of position resolution
is at least partly compensated by the $\sim 100$ times higher event rate that
can be handled by the EMU system, resulting in data of high statistical quality.
Moreover, in DR studies of polyatomic molecular ions involving heavier fragments,
also the smaller minimum distance between two fragments that can be resolved by
the EMU detector as compared to MCP/CCD-based 2D-imaging systems can be an additional advantage.

Comparing the results obtained in the present investigation of the DR of \DDHi\
with results of earlier measurements, where available, generally good agreement
is observed. In particular, the branching ratios of
78.0(0.4)\,\%, 13.5(0.3)\,\% and 8.6(0.2)\,\% measured at
$E_d=0$~eV for the D+D+H, HD+D, and D$_2$+H channels, respectively, confirm the
results recently obtained at CRYRING~\cite{PhysRevA2008-Zhaunerchyk-D2H+} using
the transmission-grid method. The branching ratios as a function of collision
energy were measured for the first time. While the general trend of the ratio of
the combined two-body to the three-body fragmentation channel looks rather
similar to what has been observed before for H$_3^+$~\cite{PhysRevLett1995-Datz-H3+}
and H$_2$D$^+$~\cite{PhysRevA1995-Datz-H2D+}, the relative branching ratios
between the D$_2$+H and HD+D channels display an unexpectedly large isotope effect. Whereas
the ratio 2$B$(D$_2$+H)/$B$(HD+D) is expected to be 1 in the absence of any 
isotope effect, this ratio is found to be enhanced by about 25\% at 
most energies investigated, and moreover reaches a so far unexplained large 
value of 3.7(0.5) at $E_d \sim5$~eV.

\begin{acknowledgments}

HB acknowledges partial support from the German-Israeli Foundation for
Scientific Research and Development (G.I.F.) under Grant I-900-231.7/2005 and by
the European Project ITS LEIF (HRPI-CT-2005-026015).
DS acknowledges support by the Weizmann Institute of Science through the Joseph
Meyerhoff program. Support by the Max-Planck Society is acknowledged.
\end{acknowledgments}


\end{document}